\documentclass[pdftex,pre,twocolumn,groupedaddress,floatfix,showpacs]{revtex4-1} 
\usepackage[utf8]{inputenc}
\usepackage{latexsym,dcolumn,graphicx,amsmath,amssymb}
\usepackage{graphicx}

\usepackage{hyperref} 
\usepackage{url}
\begin{document}

\title{Reinforced communication and social navigation: remember your friends and remember yourself}
\author{A. Mirshahvalad}
\email{atieh.mirshahvalad@physics.umu.se}
\affiliation{Integrated Science Lab, Department of Physics, Ume{\aa} University, SE-901 87 Ume{\aa}}

\author{M. Rosvall}
\email{martin.rosvall@physics.umu.se}
\affiliation{Integrated Science Lab, Department of Physics, Ume{\aa} University, SE-901 87 Ume{\aa}}
\homepage{http://www.tp.umu.se/~rosvall}
\date{\today}
\pacs{01.20.+x, 89.75.Fb,89.65.Ef}

\begin{abstract}
In social systems, people communicate with each other and form groups based on their interests. The pattern of interactions, the network, and the
ideas that flow on the network naturally evolve together. Researchers use simple models to capture the feedback between changing network patterns and
ideas on the network, but little is understood about the role of past events in the feedback process. Here we introduce a simple agent-based model to
study the coupling between peoples' ideas and social networks, and better understand the role of history in dynamic social networks.
We measure how information about ideas can be recovered from information about network structure and, the other way around, how information about network structure can be recovered from information about ideas. We find that it is in general easier to recover ideas from the network structure than vice versa.
\end{abstract}
\maketitle

\section {Introduction}
We live in the information age; in seconds, news can spread around the world through the means of modern information technology. Still, everyday communication with family, friends, and colleagues plays a major role in how information percolates through society. When this private communication is constrained to who talks to whom, information cannot move freely and will not be shared by everybody. Consequently, the pattern of interactions, the network, affects who knows what and when. But this network is not static; rather, it changes over time and peoples' current ideas will influence how new interactions form. Inevitably, the network and the ideas that flow through the network are inseparable and evolve together. Moreover, past events, stored in peoples' interest memories and the network structure, also affect future events. However, how this memory integrates with social dynamics and what role history plays in the process is still unclear. For a better understanding, here we investigate in a simple model how information about ideas is stored in the network and how information about the network is stored in ideas.

Since we don't share all our information with everybody, but rather limit most communication to friends \cite{wasserman1994social}, networks provide good tools for studying social organization \cite{carpenter}. 
To access remote connections, we must use local information \cite{kleinberg}, and the remote connections are in practice established by the flow of information through the network \cite{friedkin-infoflow}. 
In principle, everybody is connected to everybody in society, but only indirectly \cite{milgram1967small}. 
Therefore, the simplest approach to understanding how ideas form in social systems is to use static networks and study how information flows across those networks. For example, the binary Voter model \cite{liggett1999stochastic,sood2005voter}, Sznajd's consensus model \cite{sznajd2000opinion,bernardes2002election}, and the continuous relative agreement model by Deffuant et al.\ \cite{deffuant2000mixing,amblard2004role} all belong to the class of models that study group formation in social systems based on a static underlying structure \cite{PhysRevLett.85.3536,amblard2004role}. But as groups form in networks with preexisting heterogenous structures, one question remains unanswered: Where do the communities in real-world networks \cite{girvan2002community,newman2006modularity,fortunato2009community} come from?
And models that rely on preexisting heterogenous ideas or opinions \cite{mcpherson2003birds} leave unanswered the complementary question: Where did the ideas come from?
To understand the basic question of how groups form in social networks, we must consider ideas and structure simultaneously. 

Because of the interaction between structure and ideas in society, we need dynamic networks and to allow for dynamics on the network.
Therefore, recent works on the problem of group formation have focused on co-evolving networks in which people's interactions influence the network structure, and vice versa \cite{rosvall2009reinforced,iguez2009opinion}. 
For example, Holme and Newman proposed a model in which a single parameter controls the balance between the two processes: how ideas generate connections and how connections generate ideas \cite{holme2006nonequilibrium}. They observed a continuos phase transition from a heterogeneous distribution of opinions to consensus.
Kozma and Barrat generalized Deffuant's model to adaptive networks, allowing agents with similar opinions to communicate and forcing agents with sufficiently different ideas to cut their connections \cite{kozma2008consensus}. They showed that adaptive networks facilitate group formation, and that consensus in adaptive networks is harder to achieve than in static networks. 
Nardini and Kozma showed that when agents can have several opinions at the same time, for example, in the naming game model \cite{baronchelli2007role}, they more easily form consensus in adaptive networks \cite{nardini2008s}. 
Kumpula et al.\ focused on the coupling between the network topology and the interaction intensity between agents in a model with weighted links \cite{kumpula2007emergence}, thereby highlighting the structural feedback rather than the information spreading in social networks. As they increased the coupling, they observed more and more distinct communities.
For all these models, groups emerge because people align their ideas with neighbors through communication or move toward like-minded people and away from others in the network.
Groups evolve gradually over time, and history will inevitably play an important role in the dynamics.

The intrinsic memory of the system influences the feedback between structure and ideas. Past events leave behind imprints in peoples' current interests and the friends they have, which will affect their future decisions. Therefore, if we want to understand the dynamics of the system, we must understand how past experiences are integrated in the memory of the system. Some of this information is stored in peoples' interest memories and some in the interaction pattern among people. In this way, the rich dynamics of social systems are influenced both by past events stored in the different memories of the system and by the information transfer between those memories. Here we focus on the latter and ask: To what degree can ideas be recovered from the network structure, and, the other way around, to what degree can the interaction pattern be recovered from peoples' interests? 

To investigate how much people's ideas can be recovered from the interaction pattern and how much the interaction pattern can be recovered from people's ideas,
we choose an agent-based model framework with agents that can store and transmit information.
With agents that gather information by communicating with each other and store this information in an interest memory, we have shown that global knowledge can emerge from local communication \cite{rosvall2009reinforced}. 
The interest memory, which represents the agents' priorities, naturally brings history into the model framework.
To illustrate, assume a scenario in which a potential buyer uses his friendship network to look for a good deal on a car.
After gathering information through his network --- the buyer talks to his car-enthusiast friend, who talks to a car dealer friend with the perfect car in stock --- he shows up at the dealer. But the dealer is no longer willing to sell the car at the agreed price. The buyer's car-enthusiast friend and the dealer are in conflict, and the car is now reserved for another buyer.
Obviously, in this scenario the buyer's decision to visit the dealer is based on out-dated information. Over time, through repeated local communication events, the buyer has built and stored in his memory an interest in the dealer. But the interest memory is a reflection of past states of the network rather than the current state of the network, and the buyer makes a mistake.
From this example, we see that, to better understand the role of history in social dynamics, a minimal model need not contain more than the network and the agents' interest memory.
Therefore, in the next section we present a minimal version of the agent-based model presented in ref.~\cite{rosvall2009reinforced}.

\section {Modeling Communication and social navigation}

In our previous work, we developed a model for communication and social navigation based on the notion that people use local interaction to access global information. To illustrate how local interactions give access to global information, let us consider the example of the Ph.D. student who is looking for a postdoc position \cite{rosvall2009reinforced}. 
During the graduation period, the student's scientific interest becomes similar to her supervisor's interests. When she searches for 
a postdoc position she will, with high probability, go to one of her supervisor's scientific colleagues who themselves have
influenced and been influenced by the supervisor. When people navigate social networks to get better access to interesting 
information, they use information that travels across the network, propelled by communication.

When people communicate with their friends to gather information, they integrate the social system. 
The more we communicate, the more up-to-date is our information. But there is another way to get better access to up-to-date information:
People can navigate their social landscape. When we contact a friend's friend, we shortcut information pathways and approach the source of information.
In ref.~\cite{rosvall2009reinforced}, we showed that reinforced communication and social navigation generate social groups.
Social groups emerge because agents build self-organized maps of the network and navigate toward like-minded agents. The self-organized maps help agents to access information beyond their nearest neighbors. Here we show that we can, in the limit of high communication, use the shortest path as a proxy for the self-organized map and simplify the model. In the next section, we describe our simplified model framework in detail, and in Appendix \ref{app1}, we show that our simplified model can capture the same dynamics as the more complex one.  

\subsection {Model Definition}
Here we explain the main components of our model to better understand how past events stored in peoples' interests and in the network structure, and the information transfer between those memories, influence social dynamics. Our model has two main components: Agents communicate and navigate their social network to get access to information they are interested in. To implement communication and social navigation in a simple model, agents have a list of other agents' identities that represent their interest in other agents and a list of friends with whom they can communicate. When agents communicate they talk about their interests. That is, the topics of communication are restricted to the agents themselves.

We run the system with $N=100$ agents and $L=200$ links, and give each agent $a_i$ an interest memory of size $\textit{M}_i$. In principle, the agents' memories can vary in size, but for simplicity, we set the memory size to the system size.
That is, each agent $a_i$ stores $\textit{M}_i=N$ agent indices in an array that represents the interest memory. The number of times a specific agent occurs in the array reflects the relative interest in that agent.   

\begin{figure}[!ht]
\centering
\includegraphics[width=1.0 \columnwidth]{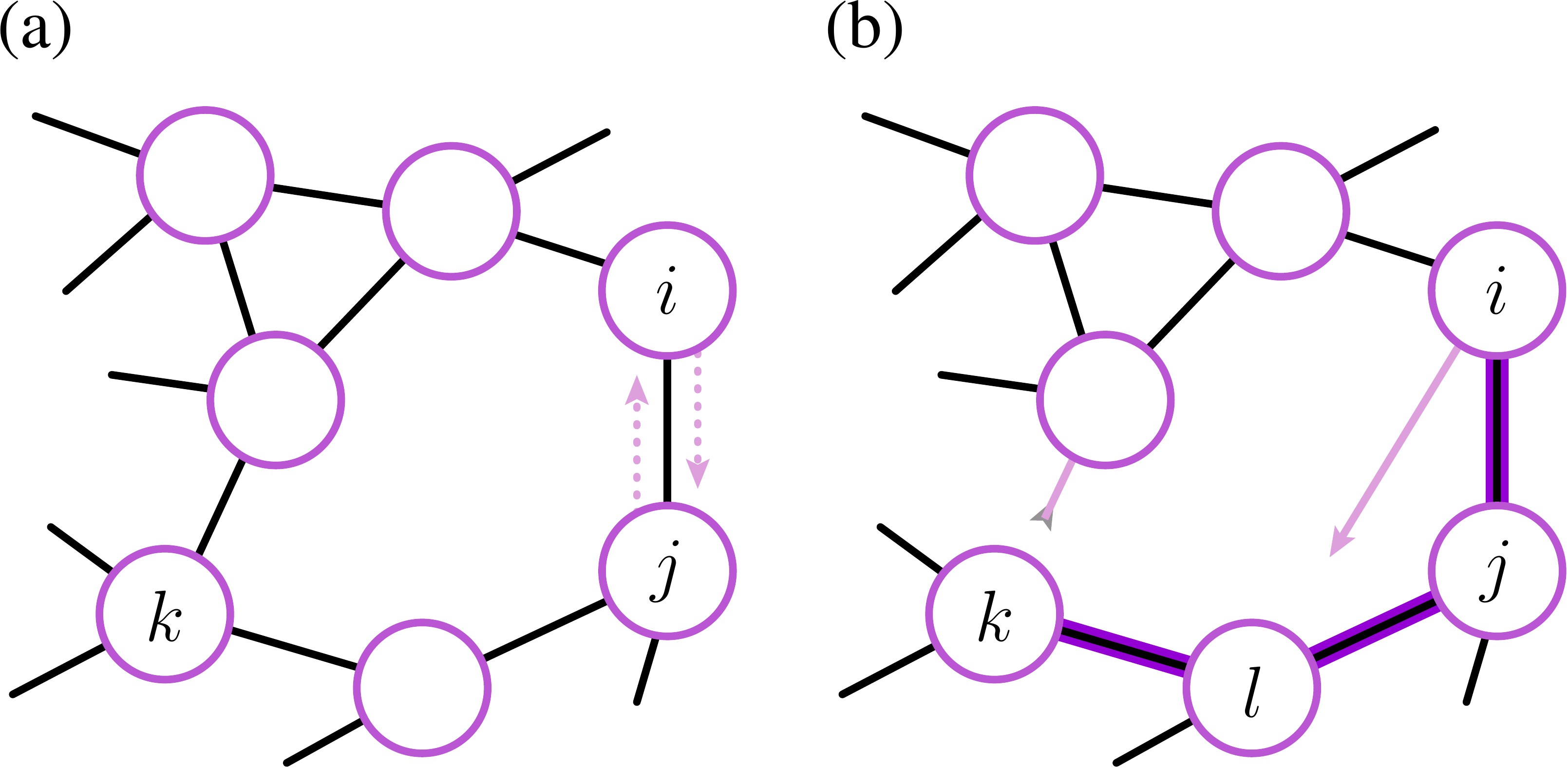}
\caption{  \label{comSocNav} Modeling communication and social navigation. \\(a) Communication: A random agent $a_i$ selects one of her neighbors $a_j$  proportional to her interest in $a_j$. In the same fashion, one of the two agents selects agent $a_k$ from her interest memory. When agent $a_i$ and $a_j$ communicate, they increase their interest in each other and also in agent $a_k$. \\(b) Social navigation: A random agent $a_i$ selects an agent $a_k$ proportional to her interest in $a_k$, and finds her friend's friend $a_l$ on the shortest path between $a_k$ and $a_i$. Agent $a_i$ forms a link to $a_l$ and shortens her distance to $a_k$. To keep the number of links fixed in the network, a random agent loses one random link.}
\end{figure}

When two agents communicate, one of the agents chooses a topic from its interest memory and both agents update their interest memories so that they become more interested in the topic and also in each other (Fig.~\ref{comSocNav}(a)).
In our previous model, the agents stored the age of the information they received and who provided the information.
When agents knew who provided the most recent information about a specific agent, they knew who to talk to to access information about this agent. 
To simplify the model, here we use the shortest path as a proxy for the self-organized map. In Appendix \ref{app1}, we show that we can recover the same dynamics as long as communication dominates over social navigation.
To shortcut the information pathways for better access to information, agents can navigate their social network and establish connections to their friends' friends (see Fig.~\ref{comSocNav}(b)).

When agents communicate, they adapt their memories to the current network, and when they navigate the social network, they seek new friends to meet their current interests. 
We control the ratio between communication and social navigation with the rewiring parameter $R$. When $R=0$, the agents only communicate and the interests fluctuate around a steady state that depends on the network configuration. Conversely, when $R=1$, the agents only rewire and the network fluctuates around a configuration that depends on the state of the agents' interests. In this paper, we study intermediate values of $R$, which generate ever-changing dynamic structures. We scale the communication-to-navigation ratio such that $L$ communication events correspond to 1 navigation step and run our simulations at $R=0.5$ if we do not mention otherwise.

In detail, in each time step, with probability $1-R$ we do $L$ communication events as follows (see Fig.~\ref{comSocNav}(a)):
 \begin{enumerate} 
 \item We pick two agents. A random agent $a_i$ with at least one link picks one of her friends $a_j$ proportional to her interest in $a_j$. 
 \item We pick the topic. Agent $a_i$ or $a_j$, chosen randomly, picks the topic $a_k$ proportional to her interest in $a_k$. 
 \item We let them exchange information. Agents $a_i$ and $a_j$ replaces two random elements in their memories with the other agent's identity and the topic $a_k$.
\end{enumerate}
With probability $R$, we do 1 social navigation step as follows (see Fig.~\ref{comSocNav}(b)):
 \begin{enumerate} 
\item We pick one agent and her target. A random agent $a_i$ picks target $a_k$ proportional to her interest in $a_k$.
\item We calculate the information route. Agent $a_l$ is agent $a_i$'s friend's friend on the shortest path between agent $a_k$ and $a_i$.  
\item We rewire. Agent $a_i$ adds a link to agent $a_l$ and a random agent loses one link.
\end {enumerate}

The rewiring ratio $R$ controls the two-way information transfer between the interest memory and the network structure. The other parameter we use is the external noise probability  $P_{\text{noise}}$. The external noise captures the concept that real-world communication is not always perfect, but is always influenced by external factors (TV, radio, the Internet, etc).
We implement the external noise with probability $P_{\text{noise}}$ by adding a random agent to an agent's interest memory instead of the current topic. As we demonstrate in Appendix \ref{app1}, external noise plays the same role as memory size in our previous model \cite{rosvall2009reinforced}: At high noise, the system forms a centralized system with hubs, and at low noise, the system self-organizes a modular structure both in interest and network structure. In our analysis below, we set the noise level to 1 percent, which generates modular structures.

\section {Results and discussion}
\begin{figure}[! ht]
\centering
\includegraphics[width=1.0 \columnwidth]{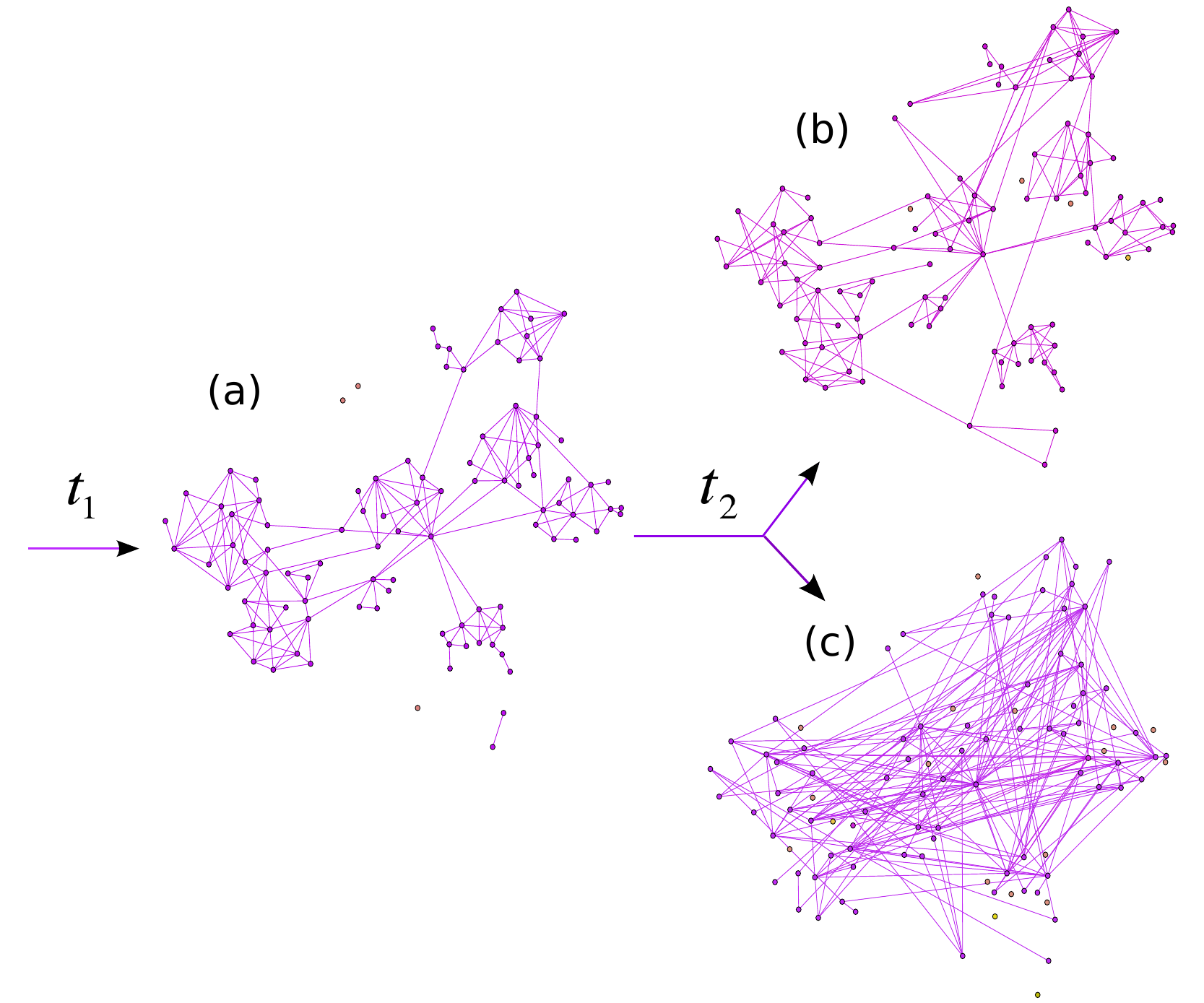}
\caption{  \label{network} Feedback between interest memory and network structure generates modular networks in which agents preserve their neighborhoods. (a) With feedback, the agents have formed a modular network at time $t_1$. (b) Continuing with feedback, the agents have preserved their neighborhoods at time $t_2$, when every link on average has been rewired one more time. (c) Continuing without feedback, the agents cannot preserve their neighborhoods and the modular structure breaks down.}
\end{figure}

Reinforced feedback between interest memory and network structure generates modular networks. We always initiate the system randomly and run the game of communication and social navigation for a long time before analyzing the outcome. Figure \ref{network}(a) shows a typical modular outcome at time $t_1$, when every link has been rewired hundreds of times. Figures \ref{network}(b) and (c) show the system at time $t_2$, when every link on average has been rewired one more time. In Fig.~\ref{network}(b) we have continued the process with feedback, and not only does the network remain modular, but agents also preserve their neighborhoods. In contrast, Fig.~\ref{network}(c) shows the outcome when we turn off the feedback at time $t_1$ and agents no longer base their interests on communication with their friends. As a result, the network looses all modular features; without feedback, agents do not preserve their neighborhoods.  

\begin{figure}[!ht]
\centering
\includegraphics[width=1.0 \columnwidth]{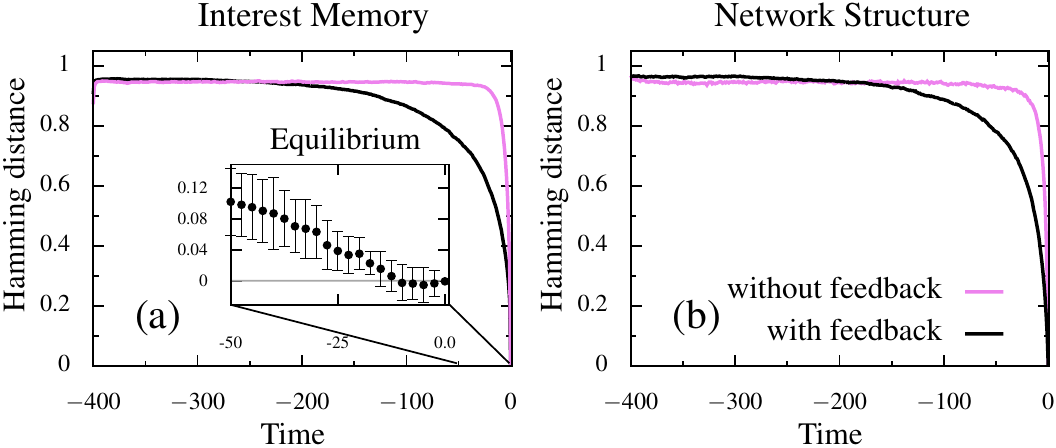}
\caption{ \label{netDyn} Feedback between interest memory and network structure slow down the dynamics of the system. 
(a) The Hamming distance of the agents' interest memories between a given time and the reference time at $t=0$, with and without feedback. The inset shows the Hamming distance at equilibrium (infinite communication) relative to the reference time out of equilibrium. (b) The dynamics of the network structure with and without feedback. In one time unit, all links have been rewired once on average.}
\end{figure}

Feedback between interest memory and network structure makes all the difference when the system evolves over time. With feedback, we see slow dynamics because agents gradually adapt themselves to the changes in the system without forgetting their past knowledge all at once. In contrast, without feedback, agents quickly adjust themselves to the changes in the system and forget their previous picture of the system. In Fig.~\ref{netDyn}, we quantify the rate of change with and without feedback between the interest memory and the network structure.
To measure memory and network distance between a given point in time and the reference point $t=0$, we tried several metrics and selected the Hamming distance for its simplicity.
The Hamming distance for interest memory measures the number of substitutions required to change an agent's interest memory from one state into another, averaged over all agents and normalized by dividing by the maximum number of substitutions.
Similarly, the Hamming distance for the network structure measures the number of substitutions necessary to change an agent's interest memory from one state into another, averaged and normalized in the same way.

Figure \ref{netDyn}(a) shows that the system evolves much more slowly with feedback than without toward a reference point at time $t=0$.
The system evolves more slowly with feedback because the current network structure affects the interest dynamics and the current interest memories affect the network dynamics.
They are two different memories about the state of the system connected by the feedback.
Because agents gather information over time, their memories do not reflect a specific network configuration, but rather they reflect the configurations of networks of the recent past.
To quantify how well the interest memory reflects past networks, in the inset of Fig.~\ref{netDyn}(a) we show the Hamming distance of memory between a given point in time at equilibrium and the reference point out of equilibrium. To reach equilibrium, we stop rewiring and let the agents communicate for a long time. 
 
The negative values from a few time steps before the reference time mean that the current interest memory in fact better captures the system of the recent past than the current state. Naturally, the  feedback also slows down the network dynamics. Figure \ref{netDyn}(b) shows that the network evolves more slowly toward the reference point with feedback than without.

Because of the feedback between memory and network, every agent is affected both by her neighbors and by her own interest memory. 
To better understand how much information is stored in the interest memory and how much is stored in the network structure, we measure how much one can be recovered from the other if one of them is scrambled.

\begin{figure}[htb]
\centering
\includegraphics[width=1.0 \columnwidth]{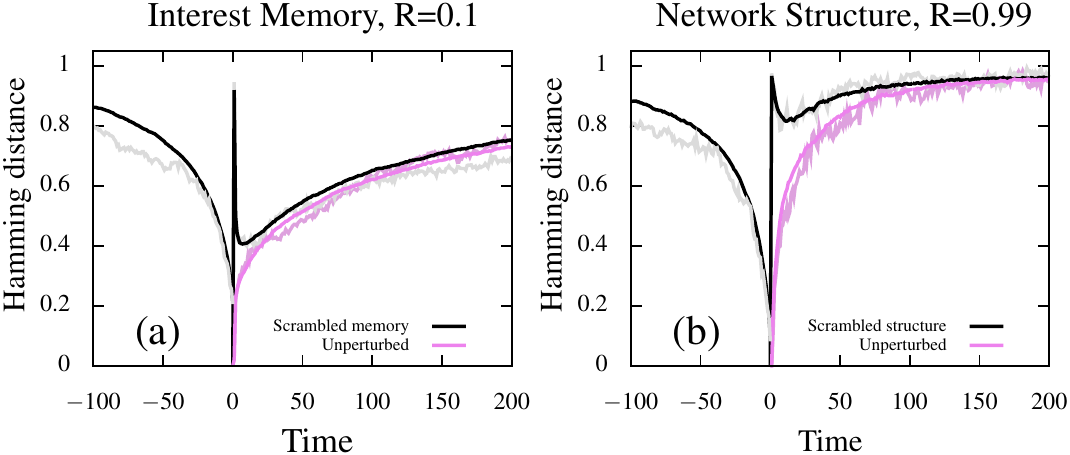}
\caption{  \label{netHamDis} System information is stored both in the agent's memory and in network structure. (a) Interest memory evolution, $R=0.1$. (b)  Network evolution, $R=0.99$ }
\end{figure}

Before scrambling the interest memories or the network structure, we run the dynamics at the default rewiring rate $R=0.5$.
At the reference time $t=0$, we scramble the memories by reshuffling the interest memories between the agents or by reshuffling the friends between the agents.
After scrambling the system, we continue with different rewiring ratios to quantify how well the system recovers.
If, after scrambling the interest memories, agents stop rewiring and only communicate with each other ($R=0$), they recover their interest memories to the equilibrium state after a few time steps. Figure \ref{netHamDis}(a) shows that the smallest rewiring rate prevents recovery to the equilibrium state. While the agents communicate to recover their interests, the social navigation based on their new interests makes them gradually diverge from the unperturbed case. The agents cannot fully recover before the network has already changed too much.

To quantify how much information the interest memories can store about the network structure, we quantify how well the network can be recovered from the interest memory.
Figure \ref{netHamDis}(b) shows how the network evolves over time.
When we scramble the network at the reference time $t=0$, we see a sharp jump in the Hamming distance of the network because agents lose their old friends and attach to random new agents. In this case, if agents stop communicating and only rewire, they reach equilibrium after a few time steps. But Fig.~\ref{netHamDis}(b) shows that 1 percent communication is enough to break the recovery. The agents find new interests while searching for their old friends and they forget their old friends. Therefore, the network dynamics diverge from the unperturbed case and recovery decreases.

\begin{figure}[htb]
\centering
\includegraphics[width=1.0\columnwidth]{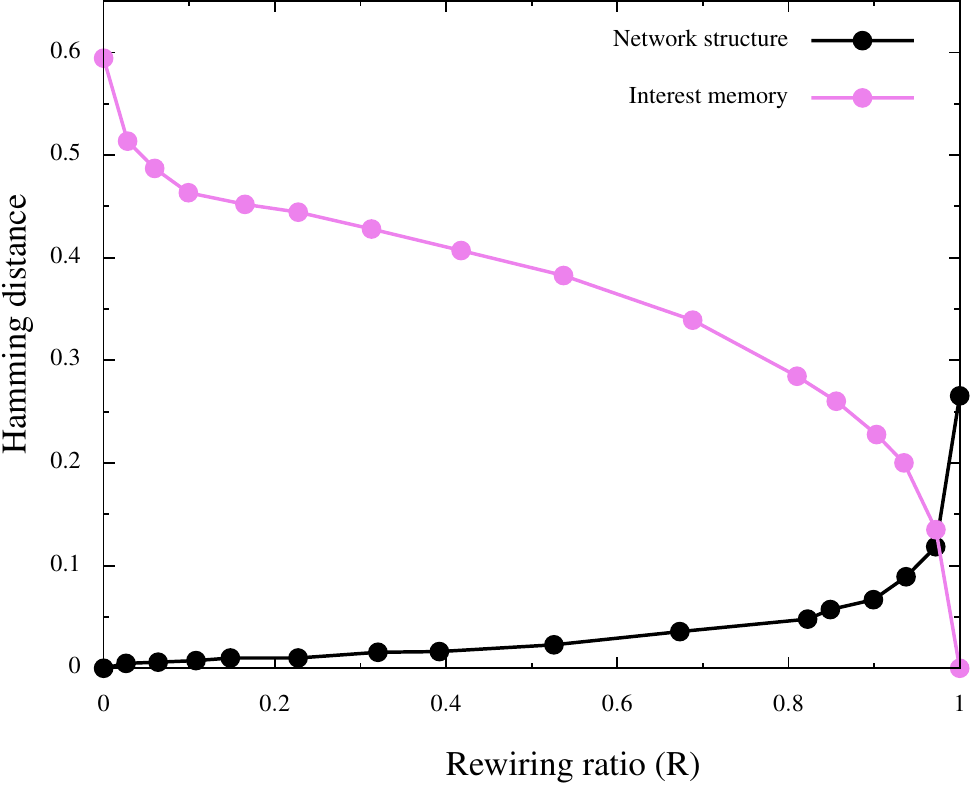}
\caption{\label{HistDep}System recovery after scrambling depends on the rewiring ratio. Low communication is necessary for network structure recovery, and a low rewiring rate is necessary for interest memory recovery.}
\end{figure}

To better understand how the feedback affects recovery at different rewiring rates, we measured the recovery for $R$ between 0 and 1. Figure \ref{HistDep} shows the results.
Very little communication is enough to significantly reduce the network recovery, because agents gradually forget their old friends when they communicate with new neighbors while searching for their old friends. An agent's typical distance to agents of interest before and after scrambling was 2.6 and 8.4, respectively, which means that after scrambling, agents are roughly 3 times farther away from their interests. It is a long way back to the old friends, and when agents communicate, they add new topics to their interest memories. Consequently, the agents lose interest in their old friends before they come back and network recovery declines. Interest memory is easier to recover, because the route back is more direct in the simpler space of the interest memory than in the more complex configuration space of the network structure. But in the same way as for network recovery, as soon as there is information transfer between the network and the interest memories, the recovery declines.

One may suggest that expanding the size of the interest memories would enhance recovery, but when we ran the simulations with memories twice as big, the results were the opposite. With bigger memories, the agents can remember longer, but the memories are also less up-to-date. At any given time, the memories are farther from equilibrium and recovery to the present network decreases. In contrast to memory recovery, the network recovery is better when agents have larger memories, because the agents lose their old friends in favor of new interests at a lower rate.  

 \section {Conclusions}
In this paper we have introduced a simple agent-based model to quantify and better understand the reinforced feedback between changing network patterns and ideas that generates social groups. We have demonstrated that history plays an important role in generating and preserving groups in the system.
In the communication and social navigation model, we have studied the information transfer between the interest memories of agents and the network structure.
To quantify how much information about the agents' ideas is stored in the network structure and, the other way around, how much information about the network structure is stored in the agents' ideas, we measured how well agents in the model could recover information stored in ideas versus information stored in the network. We conclude that the network structure contains more information about the agents' ideas than the other way around.

\appendix{}  
\begin{figure*}[!tb]
\centering
\includegraphics[width=\textwidth]{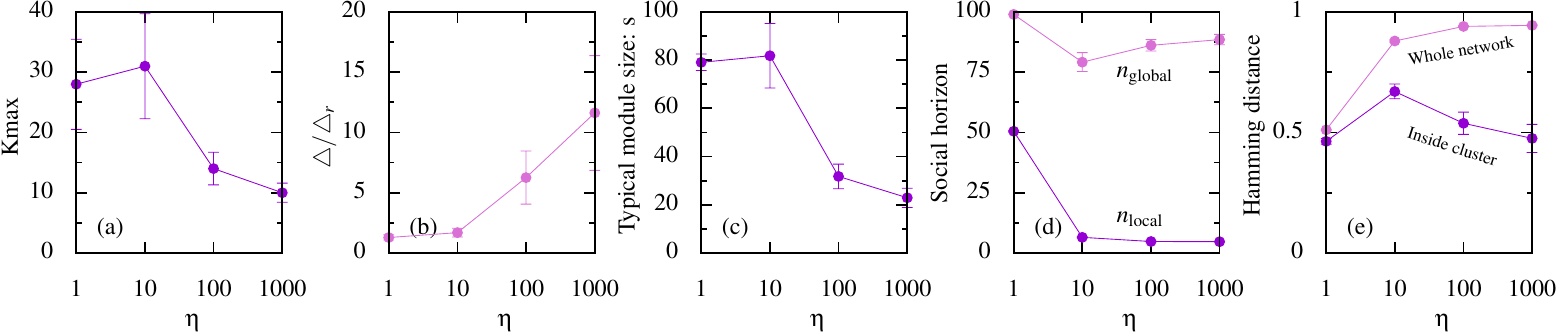} 
\caption{\label{netprop}  Reinforced communication and social navigation generate social groups. Network modularity depends on noise probability. As a function of $\eta=\frac{1}{P_{\text{noise}}}$, the panels illustrate  (a) the maximum degree in the network, (b) the abundance of triangles, (c) typical module size, (d) social horizon, and (e) Hamming distance-based measure $\Delta$.}
\end{figure*}
\section {Validating the minimal model \label{app1} }

Here we show that, as long as communication dominates over social navigation, we can capture the same dynamics in our minimal model as in our previous model \cite{rosvall2009reinforced}. Both models of communication and social navigation show how feedback between interest memories and network structure generates social groups. In the minimal model, we use the shortest paths as a proxy for communication pathways and the external noise corresponds to the memory size $\eta$ of the previous model \cite{rosvall2009reinforced}.
 
In Fig.~\ref{netprop}(a)-(d), we investigate the same four system characteristics as in ref.\ \cite{rosvall2009reinforced} and show that the results agree. The first three panels of the figure measure the structural properties of the network.
Panel (a) shows the maximum degree against the inverse noise probability, and panel (b) shows the abundance of triangles relative to random counterparts of the network. 
To quantify the modularity of the networks, we partitioned the networks \cite{rosvall2007information} into modules of sizes $\{s_l\}$ and measured the typical module size $s$, defined as the average module size that a randomly selected agent is part of,
\begin{equation}
	s=\frac{\left \langle s_l^{2}\right\rangle}{\left \langle s_l\right\rangle}.
\end{equation}
Figure \ref{netprop}(c) shows that the typical module size decreases as the external noise decreases. When the external noise decreases, the agents reinforce their interests in their friends and do not navigate away from their neighborhoods. In smaller and smaller groups, the density of links increases and the links form many triangles. The maximum degree decreases with group size, because the group size limits the number of friends an agent can have.

The transformation process from a centralized network to a modular network is also reflected in the agents' interest memories. To measure how the interest memory reveals this transformation, we measured $n_{local}$, the typical number of agents that occupy an agent's interest memory, and ${n_{global}}$, the total number of agents who receive attention from others \cite{rosvall2009reinforced}. That is, ${n_{local}}$ is the typical number of agents that an agent talks about,
\begin{equation}\label{nlocal}
	n_{local}= \left \langle \frac{N}{<m_i^{2}(j)>/<m_i(j)>} \right\rangle,
\end{equation}
and ${n_{global}}$ is the typical number of agents whom agents talk about in the entire system,
\begin{equation}\label{nglobal}
	n_{global}=\left \langle \frac{N^2}{<m^{2}(j)>/<m(j)>} \right\rangle.
\end{equation}
Figure \ref{netprop}(d) shows that, with decreasing noise level, ${n_{local}}$ decreases as the agents focus more on a few agents in more modular networks. As in the previous model, ${n_{global}}$ remains fairly constant, which means that no agents are forgotten.
The conclusion gained from Figs.~\ref{netprop}(a)-(d), is that the minimal model presented in this paper can capture the same feedback dynamics as the more complex model in ref.~\cite{rosvall2009reinforced}. 

Because it is interesting to capture the entanglement between the network structure and the interest memories, finally we investigate how much we can learn about the interest memory by looking at the network structure. We define the average Hamming distance of memory $\Delta$ as the necessary number of edits between two agents' memories, normalized by the maximum possible number of edits,
\begin{equation}
	\Delta=\left \langle\frac{\# \textit{ edits }_{i\rightarrow j}}{\textit{ max } \# \textit{ possible edits }}\right\rangle.
\end{equation} 
Figure \ref{netprop}(e) shows, with decreased external noise to the right, the Hamming distance of memory $\Delta$ averaged over agents in the same module or in the entire network. For modular networks with low noise, two agents in the same module share half their interests and almost nothing with agents in other modules.

\begin{acknowledgments}
We are grateful to Kim Sneppen and Deborah Kolp for many valuable suggestions. MR was supported by the Swedish Research Council grant 2009-5344.
\end{acknowledgments}
			


\begin{thebibliography}{24}
\expandafter\ifx\csname natexlab\endcsname\relax\def\natexlab#1{#1}\fi
\expandafter\ifx\csname bibnamefont\endcsname\relax
  \def\bibnamefont#1{#1}\fi
\expandafter\ifx\csname bibfnamefont\endcsname\relax
  \def\bibfnamefont#1{#1}\fi
\expandafter\ifx\csname citenamefont\endcsname\relax
  \def\citenamefont#1{#1}\fi
\expandafter\ifx\csname url\endcsname\relax
  \def\url#1{\texttt{#1}}\fi
\expandafter\ifx\csname urlprefix\endcsname\relax\def\urlprefix{URL }\fi
\providecommand{\bibinfo}[2]{#2}
\providecommand{\eprint}[2][]{\url{#2}}

\bibitem[{\citenamefont{Wasserman and Faust}(1994)}]{wasserman1994social}
\bibinfo{author}{\bibfnamefont{S.}~\bibnamefont{Wasserman}} \bibnamefont{and}
  \bibinfo{author}{\bibfnamefont{K.}~\bibnamefont{Faust}},
  \emph{\bibinfo{title}{{Social network analysis: Methods and applications}}}
  (\bibinfo{publisher}{Cambridge Univ Pr}, \bibinfo{year}{1994}).

\bibitem[{\citenamefont{Carpenter et~al.}(2003)\citenamefont{Carpenter,
  Esterling, and Lazer}}]{carpenter}
\bibinfo{author}{\bibfnamefont{D.}~\bibnamefont{Carpenter}},
  \bibinfo{author}{\bibfnamefont{K.}~\bibnamefont{Esterling}},
  \bibnamefont{and} \bibinfo{author}{\bibfnamefont{D.}~\bibnamefont{Lazer}},
  \bibinfo{journal}{Ration. Soc.} \textbf{\bibinfo{volume}{15}},
  \bibinfo{pages}{411} (\bibinfo{year}{2003}).

\bibitem[{\citenamefont{Kleinberg}(2000)}]{kleinberg}
\bibinfo{author}{\bibfnamefont{J.~M.} \bibnamefont{Kleinberg}},
  \bibinfo{journal}{Nature} \textbf{\bibinfo{volume}{406}},
  \bibinfo{pages}{845} (\bibinfo{year}{2000}).

\bibitem[{\citenamefont{Friedkin}(1982)}]{friedkin-infoflow}
\bibinfo{author}{\bibfnamefont{N.~E.} \bibnamefont{Friedkin}},
  \bibinfo{journal}{Soc. Networks} \textbf{\bibinfo{volume}{3}},
  \bibinfo{pages}{273} (\bibinfo{year}{1982}).

\bibitem[{\citenamefont{Milgram}(1967)}]{milgram1967small}
\bibinfo{author}{\bibfnamefont{S.}~\bibnamefont{Milgram}},
  \bibinfo{journal}{Psychology Today} \textbf{\bibinfo{volume}{2}},
  \bibinfo{pages}{60} (\bibinfo{year}{1967}).

\bibitem[{\citenamefont{Liggett}(1999)}]{liggett1999stochastic}
\bibinfo{author}{\bibfnamefont{T.}~\bibnamefont{Liggett}},
  \emph{\bibinfo{title}{{Stochastic interacting systems: contact, voter, and
  exclusion processes}}} (\bibinfo{publisher}{Springer Verlag},
  \bibinfo{year}{1999}).

\bibitem[{\citenamefont{Sood and Redner}(2005)}]{sood2005voter}
\bibinfo{author}{\bibfnamefont{V.}~\bibnamefont{Sood}} \bibnamefont{and}
  \bibinfo{author}{\bibfnamefont{S.}~\bibnamefont{Redner}},
  \bibinfo{journal}{Physical review letters} \textbf{\bibinfo{volume}{94}},
  \bibinfo{pages}{178701} (\bibinfo{year}{2005}).

\bibitem[{\citenamefont{Sznajd-Weron and Sznajd}(2000)}]{sznajd2000opinion}
\bibinfo{author}{\bibfnamefont{K.}~\bibnamefont{Sznajd-Weron}}
  \bibnamefont{and} \bibinfo{author}{\bibfnamefont{J.}~\bibnamefont{Sznajd}},
  \bibinfo{journal}{International Journal of Modern Physics C}
  \textbf{\bibinfo{volume}{11}}, \bibinfo{pages}{1157} (\bibinfo{year}{2000}).

\bibitem[{\citenamefont{Bernardes et~al.}(2002)\citenamefont{Bernardes,
  Stauffer, and Kert{\'e}sz}}]{bernardes2002election}
\bibinfo{author}{\bibfnamefont{A.}~\bibnamefont{Bernardes}},
  \bibinfo{author}{\bibfnamefont{D.}~\bibnamefont{Stauffer}}, \bibnamefont{and}
  \bibinfo{author}{\bibfnamefont{J.}~\bibnamefont{Kert{\'e}sz}},
  \bibinfo{journal}{The European Physical Journal B}
  \textbf{\bibinfo{volume}{25}}, \bibinfo{pages}{123} (\bibinfo{year}{2002}).

\bibitem[{\citenamefont{Deffuant et~al.}(2000)\citenamefont{Deffuant, Neau,
  Amblard, and Weisbuch}}]{deffuant2000mixing}
\bibinfo{author}{\bibfnamefont{G.}~\bibnamefont{Deffuant}},
  \bibinfo{author}{\bibfnamefont{D.}~\bibnamefont{Neau}},
  \bibinfo{author}{\bibfnamefont{F.}~\bibnamefont{Amblard}}, \bibnamefont{and}
  \bibinfo{author}{\bibfnamefont{G.}~\bibnamefont{Weisbuch}},
  \bibinfo{journal}{Advances in Complex Systems} \textbf{\bibinfo{volume}{3}},
  \bibinfo{pages}{87} (\bibinfo{year}{2000}).

\bibitem[{\citenamefont{Amblard and Deffuant}(2004)}]{amblard2004role}
\bibinfo{author}{\bibfnamefont{F.}~\bibnamefont{Amblard}} \bibnamefont{and}
  \bibinfo{author}{\bibfnamefont{G.}~\bibnamefont{Deffuant}},
  \bibinfo{journal}{Physica A: Statistical Mechanics and its Applications}
  \textbf{\bibinfo{volume}{343}}, \bibinfo{pages}{725} (\bibinfo{year}{2004}).

\bibitem[{\citenamefont{Castellano et~al.}(2000)\citenamefont{Castellano,
  Marsili, and Vespignani}}]{PhysRevLett.85.3536}
\bibinfo{author}{\bibfnamefont{C.}~\bibnamefont{Castellano}},
  \bibinfo{author}{\bibfnamefont{M.}~\bibnamefont{Marsili}}, \bibnamefont{and}
  \bibinfo{author}{\bibfnamefont{A.}~\bibnamefont{Vespignani}},
  \bibinfo{journal}{Phys. Rev. Lett.} \textbf{\bibinfo{volume}{85}},
  \bibinfo{pages}{3536} (\bibinfo{year}{2000}).

\bibitem[{\citenamefont{Girvan and Newman}(2002)}]{girvan2002community}
\bibinfo{author}{\bibfnamefont{M.}~\bibnamefont{Girvan}} \bibnamefont{and}
  \bibinfo{author}{\bibfnamefont{M.}~\bibnamefont{Newman}},
  \bibinfo{journal}{Proceedings of the National Academy of Sciences}
  \textbf{\bibinfo{volume}{99}}, \bibinfo{pages}{7821} (\bibinfo{year}{2002}).

\bibitem[{\citenamefont{Newman}(2006)}]{newman2006modularity}
\bibinfo{author}{\bibfnamefont{M.}~\bibnamefont{Newman}},
  \bibinfo{journal}{Proceedings of the National Academy of Sciences}
  \textbf{\bibinfo{volume}{103}}, \bibinfo{pages}{8577} (\bibinfo{year}{2006}).

\bibitem[{\citenamefont{Fortunato}(2009)}]{fortunato2009community}
\bibinfo{author}{\bibfnamefont{S.}~\bibnamefont{Fortunato}},
  \bibinfo{journal}{Physics Reports}  (\bibinfo{year}{2009}).

\bibitem[{\citenamefont{McPherson et~al.}(2003)\citenamefont{McPherson,
  Smith-Lovin, and Cook}}]{mcpherson2003birds}
\bibinfo{author}{\bibfnamefont{M.}~\bibnamefont{McPherson}},
  \bibinfo{author}{\bibfnamefont{L.}~\bibnamefont{Smith-Lovin}},
  \bibnamefont{and} \bibinfo{author}{\bibfnamefont{J.}~\bibnamefont{Cook}},
  \bibinfo{journal}{Annu. Rev. Sociol.} \textbf{\bibinfo{volume}{27}},
  \bibinfo{pages}{415} (\bibinfo{year}{2003}).

\bibitem[{\citenamefont{Rosvall and Sneppen}(2009)}]{rosvall2009reinforced}
\bibinfo{author}{\bibfnamefont{M.}~\bibnamefont{Rosvall}} \bibnamefont{and}
  \bibinfo{author}{\bibfnamefont{K.}~\bibnamefont{Sneppen}},
  \bibinfo{journal}{Physical Review E} \textbf{\bibinfo{volume}{79}},
  \bibinfo{pages}{26111} (\bibinfo{year}{2009}).

\bibitem[{\citenamefont{I{\~n}iguez et~al.}(2009)\citenamefont{I{\~n}iguez,
  Kert{\'e}sz, Kaski, and Barrio}}]{iguez2009opinion}
\bibinfo{author}{\bibfnamefont{G.}~\bibnamefont{I{\~n}iguez}},
  \bibinfo{author}{\bibfnamefont{J.}~\bibnamefont{Kert{\'e}sz}},
  \bibinfo{author}{\bibfnamefont{K.}~\bibnamefont{Kaski}}, \bibnamefont{and}
  \bibinfo{author}{\bibfnamefont{R.}~\bibnamefont{Barrio}},
  \bibinfo{journal}{Physical Review E} \textbf{\bibinfo{volume}{80}},
  \bibinfo{pages}{66119} (\bibinfo{year}{2009}).

\bibitem[{\citenamefont{Holme and Newman}(2006)}]{holme2006nonequilibrium}
\bibinfo{author}{\bibfnamefont{P.}~\bibnamefont{Holme}} \bibnamefont{and}
  \bibinfo{author}{\bibfnamefont{M.}~\bibnamefont{Newman}},
  \bibinfo{journal}{Physical Review E} \textbf{\bibinfo{volume}{74}},
  \bibinfo{pages}{56108} (\bibinfo{year}{2006}).

\bibitem[{\citenamefont{Kozma and Barrat}(2008)}]{kozma2008consensus}
\bibinfo{author}{\bibfnamefont{B.}~\bibnamefont{Kozma}} \bibnamefont{and}
  \bibinfo{author}{\bibfnamefont{A.}~\bibnamefont{Barrat}},
  \bibinfo{journal}{Physical Review E} \textbf{\bibinfo{volume}{77}},
  \bibinfo{pages}{16102} (\bibinfo{year}{2008}).

\bibitem[{\citenamefont{Baronchelli et~al.}(2007)\citenamefont{Baronchelli,
  Dall'Asta, Barrat, and Loreto}}]{baronchelli2007role}
\bibinfo{author}{\bibfnamefont{A.}~\bibnamefont{Baronchelli}},
  \bibinfo{author}{\bibfnamefont{L.}~\bibnamefont{Dall'Asta}},
  \bibinfo{author}{\bibfnamefont{A.}~\bibnamefont{Barrat}}, \bibnamefont{and}
  \bibinfo{author}{\bibfnamefont{V.}~\bibnamefont{Loreto}},
  \bibinfo{journal}{The European Physical Journal-Special Topics}
  \textbf{\bibinfo{volume}{143}}, \bibinfo{pages}{233} (\bibinfo{year}{2007}).

\bibitem[{\citenamefont{Nardini et~al.}(2008)\citenamefont{Nardini, Kozma, and
  Barrat}}]{nardini2008s}
\bibinfo{author}{\bibfnamefont{C.}~\bibnamefont{Nardini}},
  \bibinfo{author}{\bibfnamefont{B.}~\bibnamefont{Kozma}}, \bibnamefont{and}
  \bibinfo{author}{\bibfnamefont{A.}~\bibnamefont{Barrat}},
  \bibinfo{journal}{Physical Review Letters} \textbf{\bibinfo{volume}{100}},
  \bibinfo{pages}{158701} (\bibinfo{year}{2008}).

\bibitem[{\citenamefont{Kumpula et~al.}(2007)\citenamefont{Kumpula, Onnela,
  Saram{\\"a}ki, Kaski, and Kert{\'e}sz}}]{kumpula2007emergence}
\bibinfo{author}{\bibfnamefont{J.}~\bibnamefont{Kumpula}},
  \bibinfo{author}{\bibfnamefont{J.}~\bibnamefont{Onnela}},
  \bibinfo{author}{\bibfnamefont{J.}~\bibnamefont{Saram{\\"a}ki}},
  \bibinfo{author}{\bibfnamefont{K.}~\bibnamefont{Kaski}}, \bibnamefont{and}
  \bibinfo{author}{\bibfnamefont{J.}~\bibnamefont{Kert{\'e}sz}},
  \bibinfo{journal}{Physical Review Letters} \textbf{\bibinfo{volume}{99}},
  \bibinfo{pages}{228701} (\bibinfo{year}{2007}).

\bibitem[{\citenamefont{Rosvall and Bergstrom}(2007)}]{rosvall2007information}
\bibinfo{author}{\bibfnamefont{M.}~\bibnamefont{Rosvall}} \bibnamefont{and}
  \bibinfo{author}{\bibfnamefont{C.}~\bibnamefont{Bergstrom}},
  \bibinfo{journal}{Proceedings of the National Academy of Sciences}
  \textbf{\bibinfo{volume}{104}}, \bibinfo{pages}{7327} (\bibinfo{year}{2007}).

\end{thebibliography}
\end{document}